\documentclass[10pt]{article}

\usepackage{graphicx}
\usepackage{subcaption}

\usepackage[a4paper, left=35mm,right=35mm,top=34mm,bottom=34mm]{geometry}
\usepackage[utf8]{inputenc}
\usepackage[T1]{fontenc}
\usepackage[english]{babel}

\usepackage{enumerate}
\usepackage{graphicx}
\usepackage{hyperref}
\hypersetup{
    colorlinks=true,
    linkcolor=blue,
    filecolor=magenta,
    urlcolor=cyan,
}

\usepackage{mathtools,amsthm,amssymb,amsfonts}
\usepackage{algorithm}
\usepackage{algpseudocode}
\makeatletter
\def\thm@space@setup{
  \thm@preskip=10pt \thm@postskip=10pt
}
\makeatother
\theoremstyle{plain}

\theoremstyle{plain}

\theoremstyle{definition}

\theoremstyle{definition}

\theoremstyle{remark}

\theoremstyle{remark}

\usepackage{caption}
\usepackage{booktabs}
\usepackage{listings}
\usepackage{color}
\definecolor{dkgreen}{rgb}{0,0.6,0}
\definecolor{gray}{rgb}{0.5,0.5,0.5}
\definecolor{mauve}{rgb}{0.58,0,0.82}
\usepackage{enumitem}

\newcommand{\email}[1]{\protect\href{mailto:#1}{#1}}

\usepackage{xcolor}[2.11]
\colorlet{inlinkcolor}{green!50!black}
\colorlet{exlinkcolor}{red!50!black}
\if@hidelinks
\hypersetup{ hidelinks = true }
\else
\hypersetup{
  colorlinks = true,
  allcolors = inlinkcolor,
  urlcolor = exlinkcolor,
}
\fi

\newenvironment{@abssec}[1]{
        \vspace{.05in}\parindent .0in
        {\upshape\bfseries #1. }\ignorespaces
    }
    {\par\vspace{.1in}}
\renewenvironment{abstract}{\begin{@abssec}{\abstractname}}{\end{@abssec}}
\newenvironment{keywords}{\begin{@abssec}{Keywords}}{\end{@abssec}}

\usepackage{fancyhdr}

\author{
  {\normalsize Mengchen Wang }\thanks{CentraleSup\'elec, Universit\'e Paris-Saclay, 3 rue Joliot Curie, 91190 Gif-sur-Yvette, France
  (\email{frederic.magoules@hotmail.com})}  \thanks{VENISE Team, LIMSI-CNRS, Universit\'e Paris-Sud, Universit\'e Paris-Saclay, Orsay, France
  (\email{firstname.lastname@limsi.fr})}
  \and
  {\normalsize Nicolas F\'erey\footnotemark[2]}
  \and
  {\normalsize Patrick Bourdot\footnotemark[2]}
  \and
  {\normalsize Fr\'ed\'eric Magoul\`es\footnotemark[1]}
}
\title{Using asynchronous simulation approach for interactive simulation}
\date{}

\begin{document}
\maketitle
\thispagestyle{fancy}

\begin{abstract}
This paper discusses about the advantage of using asynchronous simulation in the case of interactive simulation in which user can steer and control parameters during a simulation in progress. Asynchronous models allow to compute each iteration faster to address the issues of performance needed in an highly interactive context, and our hypothesis is that get partial results faster is better than getting synchronized and final results to take a decision, in a interactive simulation context.
\end{abstract}

\begin{keywords}
Asynchronous simulation approach; interactive simulation; Distributed computing
\end{keywords}

\section{Introduction}
With the development of parallel computers, specific algorithms are designed and used to optimize the use of numerous processors at the same time and drastically increase simulation performance. In the design of these algorithms, the load balancing is very important. Load balancing means that the work must be equally distributed for all processors. When it comes to the synchronous iterations, if some processors finish the calculation faster than others, these processors must wait all other processors to finish their task. This will have much influence on the performance.

The asynchronous iterations can avoid the waiting time. The faster processors do not wait other processors and can continue the calculation at any time. There is no need to distribute equal works for every processor. In case of a physical problem, we do not need to divide the geometric domain in same sizes. This makes it much easier to make sub-domains or for unstructured meshes or distribute different types of tasks.

Interactive simulation is an approach that allows a user to see and steer a simulation in progress, using advanced interaction and rendering tools connected to a server computing the simulation. In the case of interactive simulation
architecture, one computer usually performs the simulation computation and another one gets simulation results to render it using  visualization tools.

We propose in this work to study if it possible to take advantage of asynchronous simulation models to address the computational bottleneck usually on the simulation side in interactive simulation context.


\section{Asynchronous simulation approach}

\subsection{Asynchronous interactions}
\begin{figure}
\centering
  \includegraphics[width=4in]{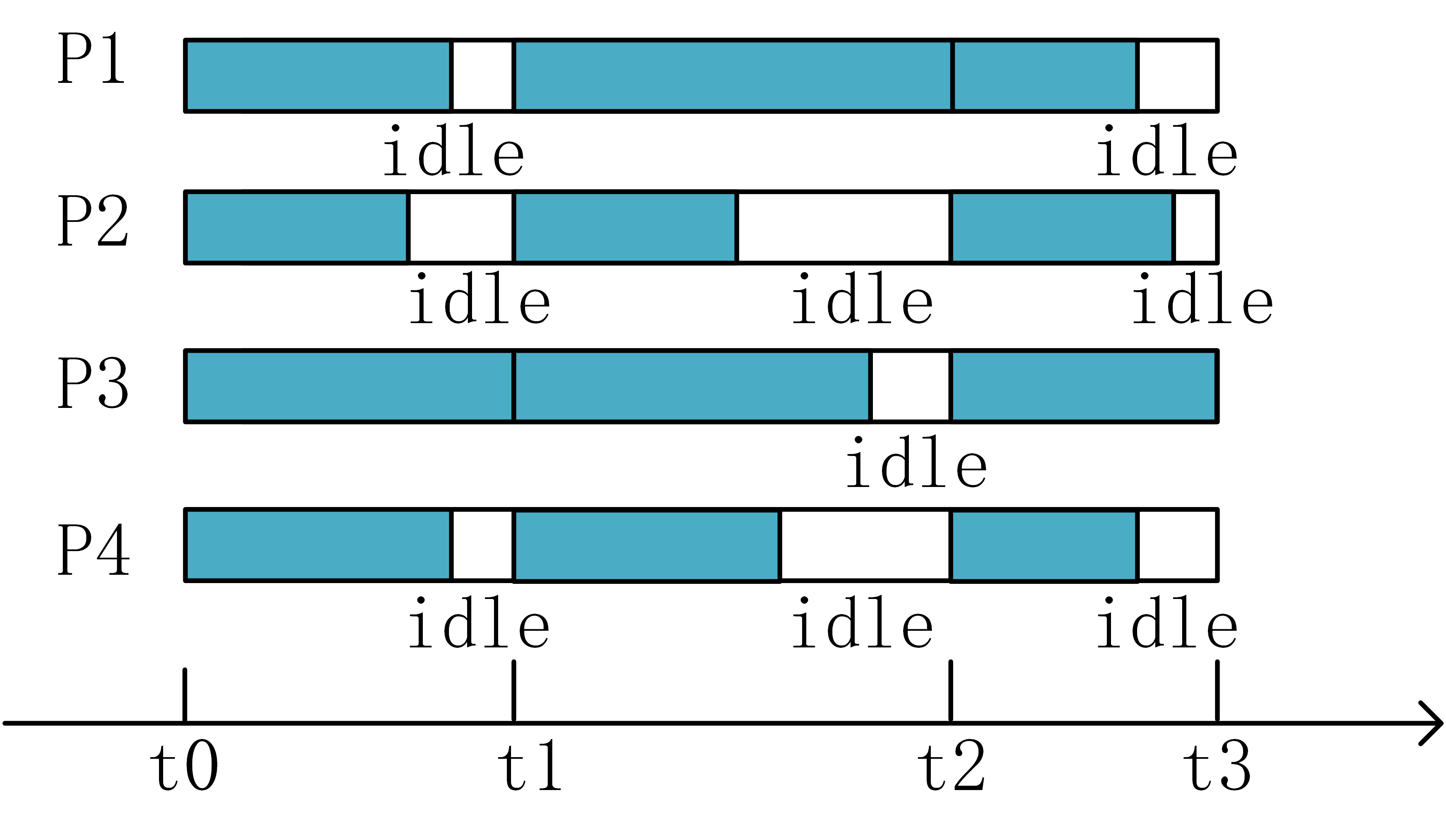}

\caption{Synchronous iterations}
\label{fig:1}       
\end{figure}

\begin{figure}
\centering
  \includegraphics[width=4in]{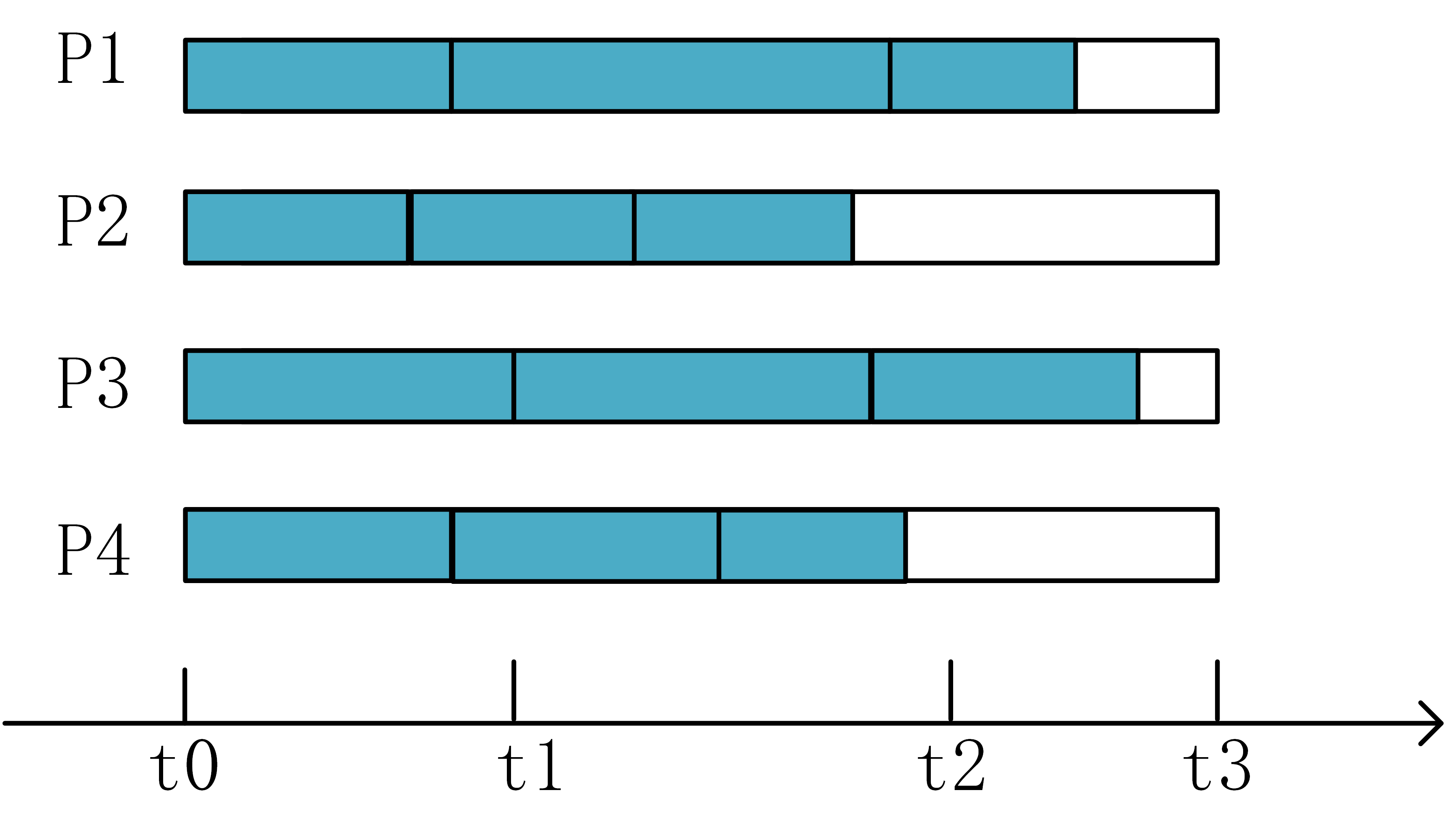}
\caption{Asynchronous iterations}
\label{fig:2}       
\end{figure}

Parallel computing is an part of computer science. It is one of the high performance solutions. Many models have been proposed for parallel computation\cite{casanova2008parallel}\cite{magoules2016parallel}. Leighton and F. Thomsonhas\cite{leighton2014introduction} have introduced the network and structures in parallel algorithms and they proposed some efficient models.
The first paper of asynchronous iterations was published in 1969 Chazan and Miranker\cite{chazan1969chaotic} to solve linear equations. It is a full overlapping of communication and computation phases during parallel iterations. This type of iteration does not require processors to wait to receive all data from other processors. In this way the processors can keep the iteration at its own pace with all the data that it has received. Asynchronous iterations are not as popular as synchronous ones as one reason maybe that people always keep in mind the load balancing. There are more and more researches on asynchronous iterations and this method is increasingly used. This numerical method feature two main interests for high performance computing (HPC).  First, the standard upper bound on the expectable acceleration factor\cite{Amdahl:1967:VSP:1465482.1465560} does not apply. Second, low sensitivity to unbalanced workload and temporary resource failures. So, asynchronous iterations are thus gaining much attention nowadays. Some experiments reported that asynchronous parallel times are much less then the synchronous parallel times. Fig.\ref{fig:1} and Fig.\ref{fig:2} shows the how asynchronous iterations can reduce communication delays to the calculation.  Bahi et al. \cite{bahi2007parallel} wrote a survey for asynchronous iterations with algorithms for solving both linear and nonlinear equations.

Recent developments have extended asynchronous iterations to parallel in time domain decomposition\cite{magoules2018asynchronous}\cite{magoules2018asynchronous1}to parallel in space domain decomposition method like sub-structuring method\cite{magoules2018asynchronous2}or like optimized Schwarz method\cite{magoules2017asynchronous}In order to monitor asynchronous iterations, stopping criteria is a real bottleneck. One example shows how we realize the domain decomposition in methods applied to European
option pricing \cite{zou2017asynchronous} and for a time-dependent case \cite{zou2017asynchronous1}.

Development in this field leads to\cite{bahi2005decentralized}\cite{el2005asynchronous} and more recently an original approach\cite{magoules2017distributed}allows to compute a residual error under asynchronous iterations with only one reduction operation.
In order to perform asynchronous simulation, a special framework should be developed like\cite{magoules:hal-01312002}\cite{Magouls2018JACK2AM} which allows easy implementation of any asynchronous iterative algorithm

The performance has been tested with  asynchronous jacobi \cite{bull2006} and shows that in some cases it requires fewer iterations with asynchronous iterations. In these cases, the possible explanation is less oscillation compared with synchronous approach. Asynchronous iteration also allows to create the progressive load balancing\cite{Zarins:2017:PLB:3149704.3149765} to reduce load imbalance without reducing stability or convergence rate.

In  iHadoop\cite{elnikety2011ihadoop}, which is a modified MapReduce model with asynchronous iterations. It is a large scale data-intensive processing framework which can schedules iterations asynchronously, without wasting bandwidth, I/O, and CPU cycles. The use of asynchronous iterations can reduce up to 81\% when compared to synchronous method. ARock\cite{peng2016arock} is a coordinate update framework using asynchronous iterations. This framework can update the shared coordinates between multiple agents(machines or processors) in a asynchronous way. In this research the ARock is compared to a sync-parallel scale. In their case, even the number of features is equally distributed to every core, the calculations are suffered from imbalanced load. The 32 cores calculation have very little speedup compared to the single core calculation, while ARock has still 75-80\% of efficiency. In the future asynchronous iteration methods maybe the methods to obtain the expected potential for the machines with thousands of processors in the future.

Implementation of asynchronous iterations requires much more than a straightforward update of classical iterations loops, depending however on the communication middleware under use. Bertsekas and Tsitsiklis\cite{Bertsekas:1989:CRT:318789.318894}  have raised an issue of the accurate evaluation of a residual based stopping criterion. JACE\cite{1271465} is a multi-threaded library aiming at the execution of asynchronous iterative algorithms based on Java Remote Method Invocation (RMI) middleware. It has been improved with a centralized volatility tolerant extension named JaceV \cite{10.1007/978-3-540-71351-7_7}. They have also used Peer-to-Peer (P2P) networks to solve large scale scientific problems \cite{4100412}. Additionally, CRAC \cite{4228296}  is another library designed to build the asynchronous applications for a grid architecture. Jack  is the first successful approach of a communication library based on MPI for synchronous and asynchronous iterations. It provides an application programming interface (API) similar to MPI routines. The difference is that the received data are saved in the reception buffer only when it is requested to be done.

Most of libraries handle the asynchronous convergence detection issue through various heuristics based on local convergence of each process, therefore they do not guarantee effective convergence state after termination. We use Jack2 to achieve asynchronous iterations. It provides a completely encapsulating API, and calculates actual global residual to realize  exact convergence testing.

\subsection{Interactive simulation}
Interactive simulation as an approach in which user can steer and control parameters during a simulation in progress. There are several advantages of using the asynchronous iteration for interactive simulation. We recall that the purpose is to get a result as fast as possible, especially for scenarios with interactive user steering.

 In the case of fluid that may take a long time, people usually wait for the result of the simulation and then analyze the result. Thus, the work has been divided to two parts. In this way, the scene cannot be interactive at all, as interactive scene requires fast calculation. The visualization in real time let us see a temporal result immediately, even though the result is not the final result, we will see the evolution of the fluid field, once we see some interesting details, we can interact with the scene, change the parameters and restart the simulation.

If we try to visualize the evolution of the calculation in synchronous iteration, this procedure will slow down the calculation because all cores must wait for the communication procedure finish to continue the calculation. That means when one core is communicating with other machines, the other CPU cores must wait until the communication ends and no calculation can be done in all cores. If the calculation has a massive size of data, the communication between machines can be very long and the calculation will be much slower. If we use this technique in the asynchronous iteration, other cores do not need to wait for the core who does the communication. The calculation can continue when the machine is transmitting data.

\section{Taking advantage of asynchronous iteration for interactive simulation : a proof of concept}
\subsection{Coupling asynchronous simulation to visualization tools}
The whole system is composed by two machines, one machine runs Unity for visualization and an other machine runs the simulation. The simulation case is a 2D heat transfer. It contains 200$\times$200 cells. We tried several different tests to see what we can benefit from the asynchronous iterations. In this test we will see that the paralyzed calculation has a bad efficiency because we exchange all data between processors. Compared with the simple calculation, the communication between processors takes much time. To achieve real-time visualization, we will always need to exchange all the data. And the communication between machines will take much longer, as explained in the next section.

\subsection{Experiments and results}
\begin{figure}
\centering
  \includegraphics[width=4in]{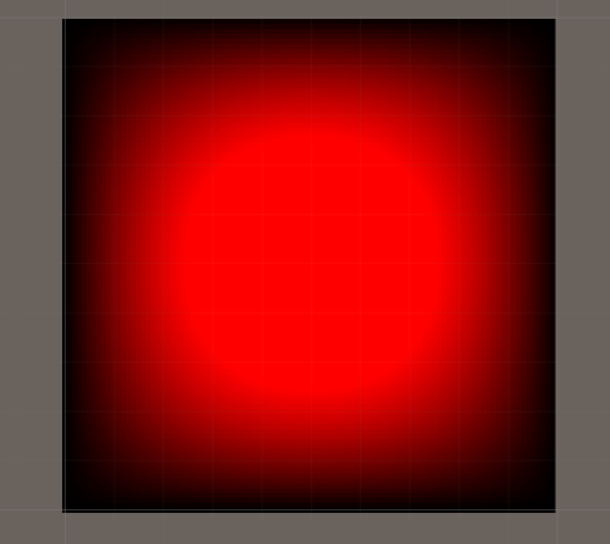}
\caption{Heat transfer visualized in Unity}
\label{fig:15}       
\end{figure}
\begin{figure}[h]
\centering
  \includegraphics[width=4in]{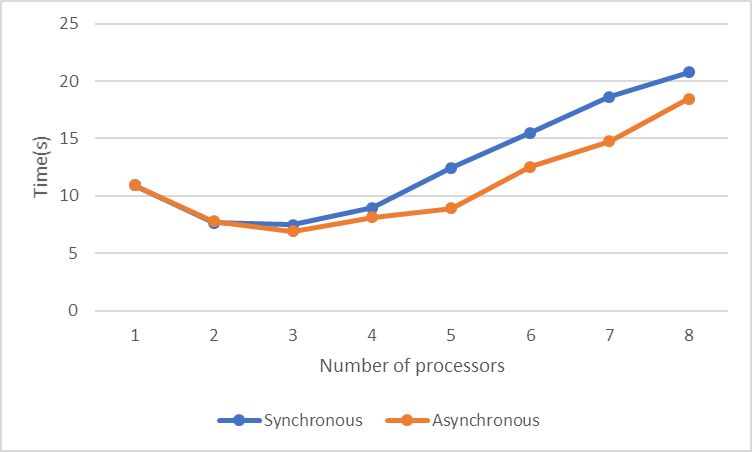}
\caption{Calculation time for a well distributed simulation}
\label{fig:5}       
\end{figure}

We tried two different tests. The first test is to see when we have a perfect balanced calculation, which means the amount of calculation on each processor are nearly the same.
The figure shows the time it takes for all the calculation to be done. We can see that even if the works are well distributed, the asynchronous iterations can improve the performance as shown in the Fig.\ref{fig:5}. We will see the parallel efficiency is very low because we run a simulation which has a simple calculation while the data amount is heavy in order to observe clearly the difference between synchronous and asynchronous iterations. The communication between processors slows down the simulation. In the case of real-time visualization, all data must be synchronized, which makes the communication between processors slower. The asynchronous iterations can reduce the impact of communication time. Also, even we well distributed the task, there will be some differences between processors.

The second test we simulated the case that the processor No 1 will deal with the communication with the machine for visualization. This test is based on a heat transfer simulation and visualization in Unity. Fig.\ref{fig:15} shows the visualization of the heat transfer when the simulation is in progress. In this case one processors will communicate with the machine which runs Unity while other processors only runs the simulation iterations.

\begin{figure}[h]
\centering
  \includegraphics[width=4in]{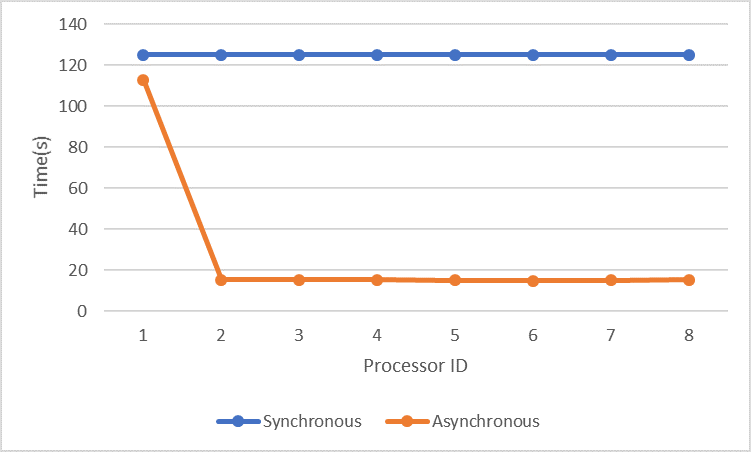}
\caption{Calculation time for each processor with one processor for communication}
\label{fig:6}       
\end{figure}

The long-distance visualization always takes time. Here we have the communication time for 12ms for each iteration. With 10k iterations, the processor 1 takes much longer than other processors. The result of the calculation time for each processor is shown in the Fig.\ref{fig:6} With synchronous iterations, all other processors will wait the processor 1 and thus all processors took more than 120 seconds to finish their calculation. While if we use the asynchronous iterations. The processors 2-7 don’t wait the processor 1. So, they all finished the calculation in around 15 seconds. This result shows us that using asynchronous iterations can avoid the impact from the communication between machines. Even if the connection between machines are unstable or slow, the calculation is always efficient. While synchronous approach will make other processors on idle and reduce the performance.

\begin{figure}[h]
\centering
  \includegraphics[width=4in]{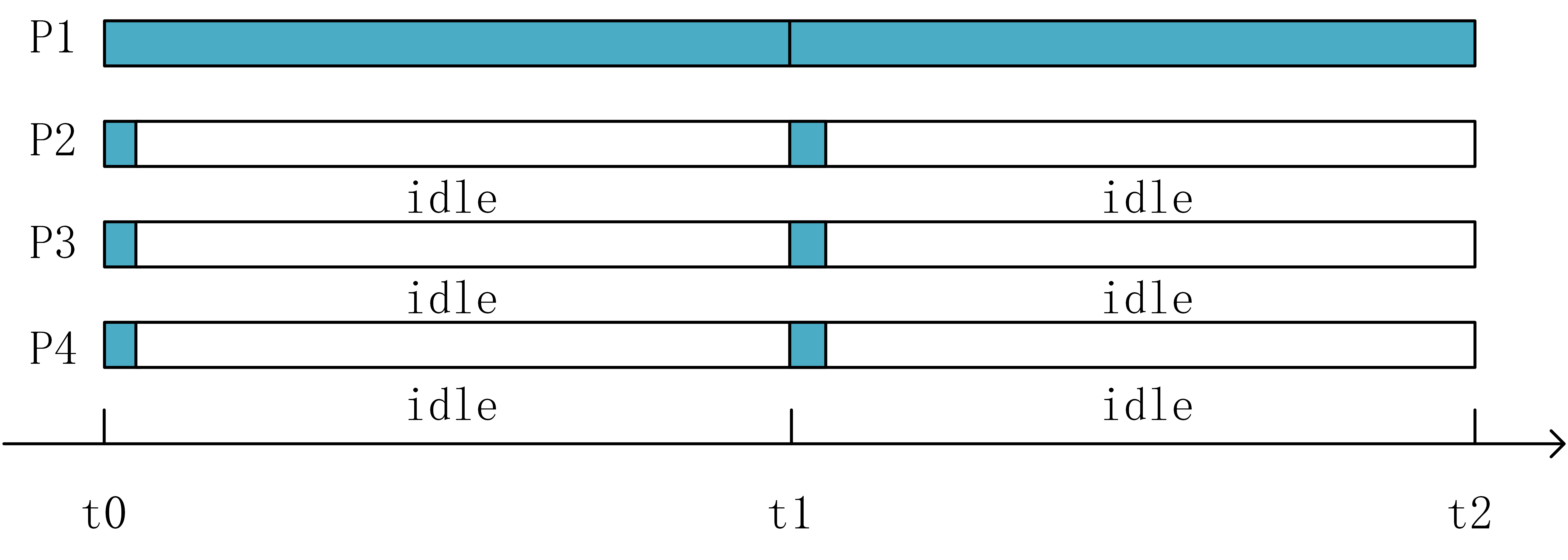}
\caption{Synchronous iterations with one processor for communication}
\label{fig:7}       
\end{figure}

 In fact, we tried a real 3D case to transfer a mesh with around 200k doubles. We use a local network with 100Mbps speed, and the time response is 1ms. It took around 100ms for data transmitting. Which means, for only 10k iterations, it will take around 15 minutes to finish, with the calculation who takes only around 1 minute. The processors will work as in the Fig.\ref{fig:7}, except the processor for the communication, other processors spend very little time on calculation and waste time on waiting. If we use a server in distance, it may take much longer for data transfer. As one processor is always transmitting data, other processors are always waiting.

Other conditions like that we have a complicated geometry, it is difficult to balance the load on every processor. For example, we have a refined area for one processor, it may be much slower compared with other processors.

\section{Conclusion}
In this paper, we described preliminary works about coupling asynchronous model to interactive simulation. In our preliminary study we show that it could be an advantage to use asynchronous models in the context of interactive simulation . Indeed, the communication between machines is usually much slower than the calculation iterations, especially for transmitting a large amount of data in distance. With synchronous iterations, other processors will wait for the communication between two machines to complete. This feature makes the task very badly distributed between processors.In our tests we can see that even with local network, the synchronous iterations are much slowed down by the processor for communication. If we use synchronous iterations, processors for calculation will carry on their work while the processor for communication is sending or receiving data. So, the asynchronous iterations are much faster than the synchronous iterations in this condition. In the future, we will use this technique to make a serious game with interactive fluid simulation.

\bibliography{paper}
\bibliographystyle{abbrv}

\end{document}